\documentclass[twocolumn,pre]{revtex4}

\usepackage{dcolumn}
\usepackage{amsmath}

\usepackage{graphicx}

\renewcommand{\O}{\mathrm{O}}

\newcommand{\eref}[1]{(\ref{#1})}
\newcommand{\etal}{{\it{}et~al.}}
\newcommand{\defn}{\textit}
\renewcommand{\vec}{\mathbf}
\newcommand{\mat}{\mathbf}

\newcommand{\Ord}{\mathop{\mathrm{O}}\nolimits}

\newlength{\figurewidth}
\begin{document}

\title{Modularity and community structure in networks}
\author{M. E. J. Newman}
\affiliation{Department of Physics and Center for the Study of Complex
Systems,\\
Randall Laboratory, University of Michigan, Ann Arbor, MI 48109--1040}
\begin{abstract}
Many networks of interest in the sciences, including a variety of social
and biological networks, are found to divide naturally into communities or
modules.  The problem of detecting and characterizing this community
structure has attracted considerable recent attention.  One of the most
sensitive detection methods is optimization of the quality function known
as ``modularity'' over the possible divisions of a network, but direct
application of this method using, for instance, simulated annealing is
computationally costly.  Here we show that the modularity can be
reformulated in terms of the eigenvectors of a new characteristic matrix
for the network, which we call the modularity matrix, and that this
reformulation leads to a spectral algorithm for community detection that
returns results of better quality than competing methods in noticeably
shorter running times.  We demonstrate the algorithm with applications to
several network data sets.
\end{abstract}
\maketitle

\subsection{Introduction}
Many systems of scientific interest can be represented as networks---sets
of nodes or \defn{vertices} joined in pairs by lines or \defn{edges}.
Examples include the Internet and the worldwide web, metabolic networks,
food webs, neural networks, communication and distribution networks, and
social networks.  The study of networked systems has a history stretching
back several centuries, but it has experienced a particular surge of
interest in the last decade, especially in the mathematical sciences,
partly as a result of the increasing availability of large-scale accurate
data describing the topology of networks in the real world.  Statistical
analyses of these data have revealed some unexpected structural features,
such as high network transitivity~\cite{WS98}, power-law degree
distributions~\cite{BA99b}, and the existence of repeated local
motifs~\cite{Milo02}; see \cite{AB02,DM02,Newman03d} for reviews.

\begin{figure}
\includegraphics[width=5cm]{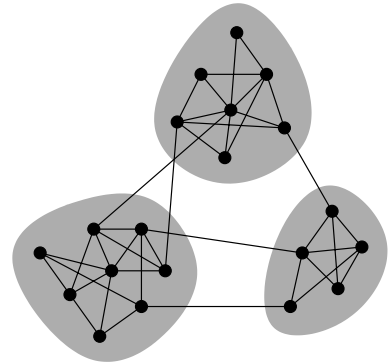}
\caption{The vertices in many networks fall naturally into groups or
communities, sets of vertices (shaded) within which there are many edges,
with only a smaller number of edges between vertices of different groups.}
\label{groups}
\end{figure}

One issue that has received a considerable amount of attention is the
detection and characterization of \defn{community structure} in
networks~\cite{Newman04b,DDDA05}, meaning the appearance of densely
connected groups of vertices, with only sparser connections between groups
(Fig.~\ref{groups}).  The ability to detect such groups could be of
significant practical importance.  For instance, groups within the
worldwide web might correspond to sets of web pages on related
topics~\cite{FLGC02}; groups within social networks might correspond to
social units or communities~\cite{GN02}.  Merely the finding that a network
contains tightly-knit groups at all can convey useful information: if a
metabolic network were divided into such groups, for instance, it could
provide evidence for a modular view of the network's dynamics, with
different groups of nodes performing different functions with some degree
of independence~\cite{HHJ03,GA05}.

Past work on methods for discovering groups in networks divides into two
principal lines of research, both with long histories.  The first, which
goes by the name of \defn{graph partitioning}, has been pursued
particularly in computer science and related fields, with applications in
parallel computing and VLSI design, among other
areas~\cite{Elsner97,Fjallstrom98}.  The second, identified by names such
as \defn{block modeling}, \defn{hierarchical clustering}, or
\defn{community structure detection}, has been pursued by sociologists and
more recently also by physicists and applied mathematicians, with
applications especially to social and biological
networks~\cite{WBB76,WF94,Newman04b}.

It is tempting to suggest that these two lines of research are really
addressing the same question, albeit by somewhat different means.  There
are, however, important differences between the goals of the two camps that
make quite different technical approaches desirable.  A typical problem in
graph partitioning is the division of a set of tasks between the processors
of a parallel computer so as to minimize the necessary amount of
interprocessor communication.  In such an application the number of
processors is usually known in advance and at least an approximate figure
for the number of tasks that each processor can handle.  Thus we know the
number and size of the groups into which the network is to be split.  Also,
the goal is usually to find the best division of the network regardless of
whether a good division even exists---there is little point in an algorithm
or method that fails to divide the network in some cases.

Community structure detection, by contrast, is perhaps best thought of as a
data analysis technique used to shed light on the structure of large-scale
network datasets, such as social networks, Internet and web data, or
biochemical networks.  Community structure methods normally assume that the
network of interest divides naturally into subgroups and the experimenter's
job is to find those groups.  The number and size of the groups is thus
determined by the network itself and not by the experimenter.  Moreover,
community structure methods may explicitly admit the possibility that no
good division of the network exists, an outcome that is itself considered
to be of interest for the light it sheds on the topology of the network.

In this paper our focus is on community structure detection in network
datasets representing real-world systems of interest.  However, both the
similarities and differences between community structure methods and graph
partitioning will motivate many of the developments that follow.

\subsection{The method of optimal modularity}
\label{modularity}
Suppose then that we are given, or discover, the structure of some network
and that we wish to determine whether there exists any natural division of
its vertices into nonoverlapping groups or communities, where these
communities may be of any size.

Let us approach this question in stages and focus initially on the problem
of whether any good division of the network exists into just two
communities.  Perhaps the most obvious way to tackle this problem is to
look for divisions of the vertices into two groups so as to minimize the
number of edges running between the groups.  This ``minimum cut'' approach
is the approach adopted, virtually without exception, in the algorithms
studied in the graph partitioning literature.  However, as discussed above,
the community structure problem differs crucially from graph partitioning
in that the sizes of the communities are not normally known in advance.  If
community sizes are unconstrained then we are, for instance, at liberty to
select the trivial division of the network that puts all the vertices in
one of our two groups and none in the other, which guarantees we will have
zero intergroup edges.  This division is, in a sense, optimal, but clearly
it does not tell us anything of any worth.  We can, if we wish,
artificially forbid this solution, but then a division that puts just one
vertex in one group and the rest in the other will often be optimal, and so
forth.

The problem is that simply counting edges is not a good way to quantify the
intuitive concept of community structure.  A good division of a network
into communities is not merely one in which there are few edges between
communities; it is one in which there are \emph{fewer than expected} edges
between communities.  If the number of edges between two groups is only
what one would expect on the basis of random chance, then few thoughtful
observers would claim this constitutes evidence of meaningful community
structure.  On the other hand, if the number of edges between groups is
significantly less than we expect by chance---or equivalently if the number
within groups is significantly more---then it is reasonable to conclude
that something interesting is going on.

This idea, that true community structure in a network corresponds to a
statistically surprising arrangement of edges, can be quantified using the
measure known as \defn{modularity}~\cite{NG04}.  The modularity is, up to a
multiplicative constant, the number of edges falling within groups minus
the expected number in an equivalent network with edges placed at random.
(A precise mathematical formulation is given below.)

The modularity can be either positive or negative, with positive values
indicating the possible presence of community structure.  Thus, one can
search for community structure precisely by looking for the divisions of a
network that have positive, and preferably large, values of the
modularity~\cite{Newman04a}.

The evidence so far suggests that this is a highly effective way to tackle
the problem.  For instance, Guimer\`a and Amaral~\cite{GA05} and later
Danon~\etal~\cite{DDDA05} optimized modularity over possible partitions of
computer-generated test networks using simulated annealing.  In direct
comparisons using standard measures, Danon~\etal\ found that this method
outperformed all other methods for community detection of which they were
aware, in most cases by an impressive margin.  On the basis of
considerations such as these we consider maximization of the modularity to
be perhaps the definitive current method of community detection, being at
the same time based on sensible statistical principles and highly effective
in practice.

Unfortunately, optimization by simulated annealing is not a workable
approach for the large network problems facing today's scientists, because
it demands too much computational effort.  A number of alternative
heuristic methods have been investigated, such as greedy
algorithms~\cite{Newman04a} and extremal optimization~\cite{DA05}.  Here we
take a different approach based on a reformulation of the modularity in
terms of the spectral properties of the network of interest.

Suppose our network contains $n$ vertices.  For a particular division of
the network into two groups let $s_i=1$ if vertex~$i$ belongs to group~1
and $s_i=-1$ if it belongs to group~2.  And let the number of edges between
vertices~$i$ and~$j$ be~$A_{ij}$, which will normally be 0 or~1, although
larger values are possible in networks where multiple edges are allowed.
(The quantities $A_{ij}$ are the elements of the so-called \defn{adjacency
matrix}.)  At the same time, the \emph{expected} number of edges between
vertices~$i$ and~$j$ if edges are placed at random is $k_ik_j/2m$, where
$k_i$ and~$k_j$ are the degrees of the vertices and $m=\frac12\sum_i k_i$
is the total number of edges in the network.  Thus the modularity can be
written
\begin{equation}
Q = {1\over4m} \sum_{ij} \biggl( A_{ij} - {k_ik_j\over2m} \biggr) s_i s_j
  = {1\over4m} \vec{s}^T\mat{B}\vec{s},
\label{qvalue}
\end{equation}
where $\vec{s}$ is the vector whose elements are the~$s_i$.  The leading
factor of $1/4m$ is merely conventional: it is included for compatibility
with the previous definition of modularity~\cite{NG04}.

We have here defined a new real symmetric matrix~$\mat{B}$ with elements
\begin{equation}
B_{ij} = A_{ij} - {k_ik_j\over2m},
\label{defsb}
\end{equation}
which we call the \defn{modularity matrix}.  Much of our attention in this
paper will be devoted to the properties of this matrix.  For the moment,
note that the elements of each of its rows and columns sum to zero, so that
it always has an eigenvector $(1,1,1,\ldots)$ with eigenvalue zero.  This
observation is reminiscent of the matrix known as the \defn{graph
Laplacian}~\cite{Chung97}, which is the basis for one of the best-known
methods of graph partitioning, \defn{spectral
partitioning}~\cite{Fiedler73,PSL90}, and has the same property.  And
indeed, the methods presented in this paper have many similarities to
spectral partitioning, although there are some crucial differences as well,
as we will see.

Given Eq.~\eref{qvalue}, we proceed by writing $\vec{s}$ as a linear
combination of the normalized eigenvectors $\vec{u}_i$ of $\mat{B}$ so that
$\vec{s} = \sum_{i=1}^n a_i \vec{u}_i$ with $a_i=\vec{u}_i^T\cdot\vec{s}$.
Then we find
\begin{equation}
Q = \sum_i a_i \vec{u}_i^T \mat{B} \sum_j a_j \vec{u}_j
  = \sum_{i=1}^n (\vec{u}_i^T\cdot\vec{s})^2 \beta_i,
\label{qai}
\end{equation}
where $\beta_i$ is the eigenvalue of $\mat{B}$ corresponding to
eigenvector~$\vec{u}_i$.  Here, and henceforth, we drop the leading factor
of $1/4m$ for the sake of brevity.

Let us assume that the eigenvalues are labeled in decreasing order,
$\beta_1\ge\beta_2\ge\ldots\ge\beta_n$.  We wish to maximize the modularity
by choosing an appropriate division of the network, or equivalently by
choosing the value of the index vector~$\vec{s}$.  This means choosing
$\vec{s}$ so as to concentrate as much weight as possible in the terms of
the sum involving the largest (most positive) eigenvalues.  If there were
no other constraints on our choice of~$\vec{s}$ (apart from normalization),
this would be an easy task: we would simply chose $\vec{s}$ proportional to
the eigenvector~$\vec{u}_1$.  This places all of the weight in the term
involving the largest eigenvalue~$\beta_1$, the other terms being
automatically zero, since the eigenvectors are orthogonal.

Unfortunately, there is another constraint on the problem imposed by the
restriction of the elements of $\vec{s}$ to the values~$\pm1$, which means
$\vec{s}$ cannot normally be chosen proportional to~$\vec{u}_1$.  This
makes the optimization problem much more difficult.  Indeed, it seems
likely that the problem is NP-hard computationally, since it is formally
equivalent to an instance of the NP-hard MAX-CUT problem.  This makes it
improbable that a simple procedure exists for finding the
optimal~$\vec{s}$, so we turn instead to approximate methods.

An optimization problem similar to this one appears in the development of
the spectral partitioning method and in that case a very simple
approximation is found to be effective, namely maximizing the term
involving the leading eigenvalue and completely ignoring all the others.
As we now show, the same approach turns out to be effective here too: we
simply ignore the inconvenient fact that it is not possible to make
$\vec{s}$ perfectly parallel to $\vec{u}_1$ and go ahead and maximize the
term in $\beta_1$ anyway.

Given that we are free to choose the sizes of our two groups of vertices,
the greatest value of the coefficient $(\vec{u}_1^T\cdot\vec{s})^2$ in this
term is achieved by dividing the vertices according to the signs of the
elements in~$\vec{u}_1$---all vertices whose corresponding elements in
$\vec{u}_1$ are positive go in one group and all the rest in the other
group.  So this is our algorithm: we compute the leading eigenvector of the
modularity matrix and divide the vertices into two groups according to the
signs of the corresponding elements in this vector.

We immediately notice some satisfying features of this method.  First, as
we have made clear, it works even though the sizes of the communities are
not specified.  Unlike conventional partitioning methods that minimize the
number of between-group edges, there is no need to constrain the group
sizes or to artificially forbid the trivial solution with all vertices in a
single group.  There is an eigenvector $(1,1,1,\ldots)$ corresponding to
such a trivial solution, but its eigenvalue is zero.  All other
eigenvectors are orthogonal to this one and hence must possess both
positive and negative elements.  Thus, so long as there is any positive
eigenvalue our method will not put all vertices in the same group.

It is however possible for there to be no positive eigenvalues of the
modularity matrix.  In this case the leading eigenvector $\emph{is}$ the
vector $(1,1,1,\ldots)$ corresponding to all vertices in a single group
together.  But this is precisely the correct result: the algorithm is in
this case telling us that there is no division of the network that results
in positive modularity, as can immediately be seen from Eq.~\eref{qai},
since all terms in the sum will be zero or negative.  The modularity of the
undivided network is zero, which is the best that can be achieved.  This is
an important feature of our algorithm.  The algorithm has the ability not
only to divide networks effectively, but also to refuse to divide them when
no good division exists.  We will call the networks in this latter case
\defn{indivisible}.  That is, a network is indivisible if the modularity
matrix has no positive eigenvalues.  This idea will play a crucial role in
later developments.

Our algorithm as we have described it makes use only of the signs of the
elements of the leading eigenvector, but the magnitudes convey information
too.  Vertices corresponding to elements of large magnitude make large
contributions to the modularity, Eq.~\eref{qai}, and conversely for small
ones.  This means that moving a vertex corresponding to an element of small
magnitude from one group to the other makes little difference to~$Q$.  In
other words, the magnitudes of the elements are a measure of how
``strongly'' a vertex belongs to one community or the other, and vertices
with elements close to zero are, in a sense, on the borderline between
communities.  Thus our algorithm allows us not merely to divide the
vertices into groups, but to place them on a continuous scale of ``how
much'' they belong to one group or the other.

\begin{figure}
\includegraphics[width=8cm]{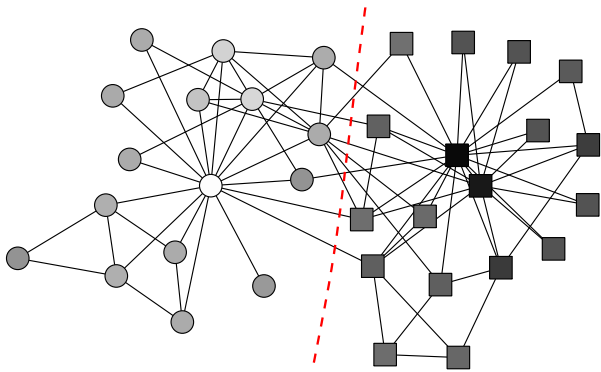}
\caption{Application of our eigenvector-based method to the ``karate club''
network of Ref.~\cite{Zachary77}.  Shapes of vertices indicate the
membership of the corresponding individuals in the two known factions of
the network while the dotted line indicates the split found by the
algorithm, which matches the factions exactly.  The shades of the vertices
indicate the strength of their membership, as measured by the value of the
corresponding element of the eigenvector.}
\label{zachary}
\end{figure}

As an example of this algorithm we show in Fig.~\ref{zachary} the result of
its application to a famous network from the social science literature,
which has become something of a standard test for community detection
algorithms.  The network is the ``karate club'' network of
Zachary~\cite{Zachary77}, which shows the pattern of friendships between
the members of a karate club at a US university in the 1970s.  This example
is of particular interest because, shortly after the observation and
construction of the network, the club in question split in two as a result
of an internal dispute.  Applying our eigenvector-based algorithm to the
network, we find the division indicated by the dotted line in the figure,
which coincides exactly with the known division of the club in real life.

The vertices in Fig.~\ref{zachary} are shaded according to the values of
the elements in the leading eigenvector of the modularity matrix, and these
values seem also to accord well with known social structure within the
club.  In particular, the three vertices with the heaviest weights, either
positive or negative (black and white vertices in the figure), correspond
to the known ringleaders of the two factions.

\subsection{Dividing networks into more than two communities}
In the preceding section we have given a simple matrix-based method for
finding a good division of a network into two parts.  Many networks,
however, contain more than two communities, so we would like to extend our
method to find good divisions of networks into larger numbers of parts.
The standard approach to this problem, and the one adopted here, is
repeated division into two: we use the algorithm of the previous section
first to divide the network into two parts, then divide those parts, and so
forth.

In doing this it is crucial to note that it is not correct, after first
dividing a network in two, to simply delete the edges falling between the
two parts and then apply the algorithm again to each subgraph.  This is
because the degrees appearing in the definition, Eq.~\eref{qvalue}, of the
modularity will change if edges are deleted, and any subsequent
maximization of modularity would thus maximize the wrong quantity.
Instead, the correct approach is to define for each subgraph~$g$ a new
$n_g\times n_g$ modularity matrix~$\mat{B}^{(g)}$, where $n_g$ is the
number of vertices in the subgraph.  The correct definition of the element
of this matrix for vertices~$i,j$ is
\begin{equation}
B^{(g)}_{ij} = A_{ij} - {k_ik_j\over2m}
               - \delta_{ij} \biggl[ k_i^{(g)} - k_i {d_g\over2m} \biggr],
\label{defsbg}
\end{equation}
where $k_i^{(g)}$ is the degree of vertex~$i$ within subgraph~$g$ and $d_g$
is the sum of the (total) degrees~$k_i$ of the vertices in the subgraph.
Then the subgraph modularity $Q_g = \vec{s}^T\mat{B}^{(g)}\vec{s}$
correctly gives the additional contribution to the total modularity made by
the division of this subgraph.  In particular, note that if the subgraph is
undivided, $Q_g$~is correctly zero.  Note also that for a complete network
Eq.~\eref{defsbg} reduces to the previous definition for the modularity
matrix, Eq.~\eref{defsb}, since $k_i^{(g)}\to k_i$ and $d_g\to2m$ in that
case.

In repeatedly subdividing our network, an important question we need to
address is at what point to halt the subdivision process.  A nice feature
of our method is that it provides a clear answer to this question: if there
exists no division of a subgraph that will increase the modularity of the
network, or equivalently that gives a positive value for~$Q_g$, then there
is nothing to be gained by dividing the subgraph and it should be left
alone; it is indivisible in the sense of the previous section.  This
happens when there are no positive eigenvalues to the
matrix~$\mat{B}^{(g)}$, and thus our leading eigenvalue provides a simple
check for the termination of the subdivision process: if the leading
eigenvalue is zero, which is the smallest value it can take, then the
subgraph is indivisible.

Note, however, that while the absence of positive eigenvalues is a
sufficient condition for indivisibility, it is not a necessary one.  In
particular, if there are only small positive eigenvalues and large negative
ones, the terms in Eq.~\eref{qai} for negative~$\beta_i$ may outweigh those
for positive.  It is straightforward to guard against this possibility,
however: we simply calculate the modularity contribution for each proposed
split directly and confirm that it is greater than zero.

Thus our algorithm is as follows.  We construct the modularity matrix for
our network and find its leading (most positive) eigenvalue and
eigenvector.  We divide the network into two parts according to the signs
of the elements of this vector, and then repeat for each of the parts.  If
at any stage we find that the proposed split makes a zero or negative
contribution to the total modularity, we leave the corresponding subgraph
undivided.  When the entire network has been decomposed into indivisible
subgraphs in this way, the algorithm ends.

One immediate corollary of this approach is that all ``communities'' in the
network are, by definition, indivisible subgraphs.  A number of authors
have in the past proposed formal definitions of what a community
is~\cite{WF94,FLGC02,Radicchi04}.  The present method provides an
alternative, first-principles definition of a community as an indivisible
subgraph.

\subsection{Further techniques for modularity maximization}
In this section we describe briefly another method we have investigated for
dividing networks in two by modularity optimization, which is entirely
different from our spectral method.  Although not of especial interest on
its own, this second method is, as we will shortly show, very effective
when \emph{combined} with the spectral method.

Let us start with some initial division of our vertices into two groups:
the most obvious choice is simply to place all vertices in one of the
groups and no vertices in the other.  Then we proceed as follows.  We find
among the vertices the one that, when moved to the other group, will give
the biggest increase in the modularity of the complete network, or the
smallest decrease if no increase is possible.  We make such moves
repeatedly, with the constraint that each vertex is moved only once.  When
all $n$ vertices have been moved, we search the set of intermediate states
occupied by the network during the operation of the algorithm to find the
state that has the greatest modularity.  Starting again from this state, we
repeat the entire process iteratively until no further improvement in the
modularity results.  Those familiar with the literature on graph
partitioning may find this algorithm reminiscent of the Kernighan--Lin
algorithm~\cite{KL70}, and indeed the Kernighan--Lin algorithm provided the
inspiration for our method.

Despite its simplicity, we find that this method works moderately well.  It
is not competitive with the best previous methods, but it gives respectable
modularity values in the trial applications we have made.  However, the
method really comes into its own when it is used \emph{in combination} with
the spectral method introduced earlier.  It is a common approach in
standard graph partitioning problems to use spectral partitioning based on
the graph Laplacian to give an initial broad division of a network into two
parts, and then refine that division using the Kernighan--Lin algorithm.
For community structure problems we find that the equivalent joint strategy
works very well.  Our spectral approach based on the leading eigenvector of
the modularity matrix gives an excellent guide to the general form that the
communities should take and this general form can then be fine-tuned by our
vertex moving method, to reach the best possible modularity value.  The
whole procedure is repeated to subdivide the network until every remaining
subgraph is indivisible, and no further improvement in the modularity is
possible.

Typically, the fine-tuning stages of the algorithm add only a few percent
to the final value of the modularity, but those few percent are enough to
make the difference between a method that is merely good and one that is,
as we will see, exceptional.

\subsection{Example applications}
\label{examples}
In practice, the algorithm developed here gives excellent results.  For a
quantitative comparison between our algorithm and others we follow Duch and
Arenas~\cite{DA05} and compare values of the modularity for a variety of
networks drawn from the literature.  Results are shown in
Table~\ref{results} for six different networks---the exact same six as used
by Duch and Arenas.  We compare modularity figures against three previously
published algorithms: the betweenness-based algorithm of Girvan and
Newman~\cite{GN02}, which is widely used and has been incorporated into
some of the more popular network analysis programs (denoted GN in the
table); the fast algorithm of Clauset~\etal~\cite{CNM04} (CNM), which
optimizes modularity using a greedy algorithm; and the extremal
optimization algorithm of Duch and Arenas~\cite{DA05} (DA), which is
arguably the best previously existing method, by standard measures, if one
discounts methods impractical for large networks, such as exhaustive
enumeration of all partitions or simulated annealing.

The table reveals some interesting patterns.  Our algorithm clearly
outperforms the methods of Girvan and Newman and of Clauset~\etal\ for all
the networks in the task of optimizing the modularity.  The extremal
optimization method on the other hand is more competitive.  For the smaller
networks, up to around a thousand vertices, there is essentially no
difference in performance between our method and extremal optimization; the
modularity values for the divisions found by the two algorithms differ by
no more than a few parts in a thousand for any given network.  For larger
networks, however, our algorithm does better than extremal optimization,
and furthermore the gap widens as network size increases, to a maximum
modularity difference of about a 6\% for the largest network studied.  For
the very large networks that have been of particular interest in the last
few years, therefore, it appears that our method for detecting community
structure may be the most effective of the methods considered here.

\begin{table}[t]
\setlength{\tabcolsep}{4pt}
\begin{tabular}{l|r|cccc}
               &           & \multicolumn{4}{c}{modularity $Q$} \\
network        &  size $n$ & GN    & CNM   & DA    & this paper \\
\hline
karate         &        34 & 0.401 & 0.381 & 0.419 & 0.419 \\
jazz musicians &       198 & 0.405 & 0.439 & 0.445 & 0.442 \\
metabolic      &       453 & 0.403 & 0.402 & 0.434 & 0.435 \\
email          &      1133 & 0.532 & 0.494 & 0.574 & 0.572 \\
key signing    & $10\,680$ & 0.816 & 0.733 & 0.846 & 0.855 \\
physicists     & $27\,519$ &    -- & 0.668 & 0.679 & 0.723
\end{tabular}
\caption{Comparison of modularities for the network divisions found by
the algorithm described here and three other previously published methods
as described in the text, for six networks of varying sizes.  The networks
are, in order, the karate club network of Zachary~\cite{Zachary77}, the
network of collaborations between early jazz musicians of Gleiser and
Danon~\cite{GD03}, a metabolic network for the nematode
\textit{C. elegans}~\cite{Jeong00}, a network of email contacts between
students~\cite{EMB02}, a trust network of mutual signing of cryptography
keys~\cite{Guardiola02}, and a coauthorship network of scientists working
on condensed matter physics~\cite{Newman01a}.  No modularity figure is
given for the last network with the GN algorithm because the slow
$\Ord(n^3)$ operation of the algorithm prevents its application to such
large systems.}
\label{results}
\end{table}

The modularity values given in Table~\ref{results} provide a useful
quantitative measure of the success of our algorithm when applied to
real-world problems.  It is worthwhile, however, also to confirm that it
returns sensible divisions of networks in practice.  We have given one
example demonstrating such a division in Fig.~\ref{zachary}.  We have also
checked our method against many of the example networks used in previous
studies~\cite{GN02,NG04}.  Here we give two more examples, both involving
network representations of US politics.

The first example is a network of books on politics, compiled by V.~Krebs
(unpublished, but see \verb|www.orgnet.com|).  In this network the vertices
represent 105 recent books on American politics bought from the on-line
bookseller Amazon.com, and edges join pairs of books that are frequently
purchased by the same buyer.  Books were divided (by the present author)
according to their stated or apparent political alignment---liberal or
conservative---except for a small number of books that were explicitly
bipartisan or centrist, or had no clear affiliation.

\begin{figure}
\includegraphics[width=\columnwidth]{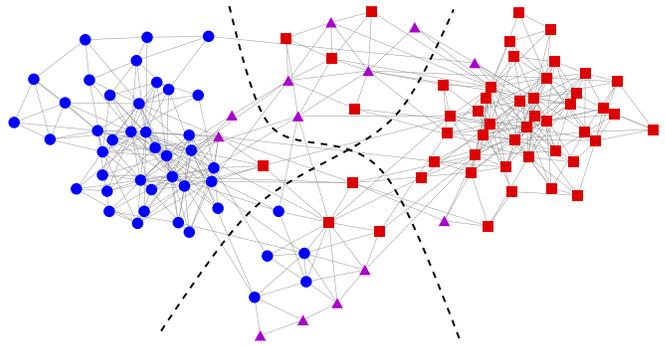}
\caption{Krebs' network of books on American politics.  Vertices represent
books and edges join books frequently purchased by the same readers.
Dotted lines divide the four communities found by our algorithm and shapes
represent the political alignment of the books: circles (blue) are liberal,
squares (red) are conservative, triangles (purple) are centrist or
unaligned.}
\label{books}
\end{figure}

Figure~\ref{books} shows the result of feeding this network through our
algorithm.  The algorithm finds four communities of vertices, denoted by
the dotted lines in the figure.  As we can see, one of these communities
consists almost entirely of liberal books and one almost entirely of
conservative books.  Most of the centrist books fall in the two remaining
communities.  Thus these books appear to form ``communities'' of
copurchasing that align closely with political views, a result that
encourages us to believe that our algorithm is capable of extracting
meaningful results from raw network data.  It is particularly interesting
to note that the centrist books belong to their own communities and are
not, in most cases, merely lumped in with the liberals or conservatives;
this may indicate that political moderates form their own community of
purchasing.

For our second example, we consider a network of political commentary web
sites, also called ``weblogs'' or ``blogs,'' compiled from on-line
directories by Adamic and Glance~\cite{AG05}, who also assigned a political
alignment, conservative or liberal, to each blog based on content.  The
1225 vertices in the network studied here correspond to the 1225 blogs in
the largest component of Adamic and Glance's network, and undirected edges
connect vertices if either of the corresponding blogs contained a hyperlink
to the other on its front page.  On feeding this network through our
algorithm we discover that the network divides cleanly into conservative
and liberal communities and, remarkably, the optimal modularity found is
for a division into just two communities.  One community has 638 vertices
of which 620 (97\%) represent conservative blogs.  The other has 587
vertices of which 548 (93\%) represent liberal blogs.  The algorithm found
no division of either of these two groups that gives any positive
contribution to the modularity; these groups are ``indivisible'' in the
sense defined in this paper.  This behavior is unique in our experience
among networks of this size and is perhaps a testament not only to the
widely noted polarization of the current political landscape in the United
States but also to the strong cohesion of the two factions.

Finally, we mention that as well as being accurate our method is also fast.
It can be shown that the running time of the algorithm scales with system
size as $\Ord(n^2\log n)$ for the typical case of a sparse graph.  This is
considerably better than the $\O(n^3)$ running time of the betweenness
algorithm~\cite{GN02}, and slightly better than the $\O(n^2\log^2 n)$ of
the extremal optimization algorithm~\cite{DA05}.  It is not as good as the
$\O(n\log^2 n)$ for the greedy algorithm of~\cite{CNM04}, but our results
are of far better quality than those for the greedy algorithm.  In
practice, running times are reasonable for networks up to about $100\,000$
vertices with current computers.  For the largest of the networks studied
here, the collaboration network, which has about $27\,000$ vertices, the
algorithm takes around 20 minutes to run on a standard personal computer.

\subsection{Conclusions}
In this paper we have examined the problem of detecting community structure
in networks, which we frame as an optimization task in which one searches
for the maximal value of the quantity known as modularity over possible
divisions of a network.  We have shown that this problem can be rewritten
in terms of the eigenvalues and eigenvectors of a matrix we call the
modularity matrix, and by exploiting this transformation we have created a
new computer algorithm for community detection that demonstrably
outperforms the best previous general-purpose algorithms in terms of both
quality of results and speed of execution.  We have applied our algorithm
to a variety of real-world network data sets, including social and
biological examples, showing it to give both intuitively reasonable
divisions of networks and quantitatively better results as measured by the
modularity.

\begin{acknowledgments}
The author would like to thank Lada Adamic, Alex Arenas, and Valdis Krebs
for providing network data and for useful comments and suggestions.  This
work was funded in part by the National Science Foundation under grant
number DMS--0234188 and by the James S. McDonnell Foundation.
\end{acknowledgments}

\end{document}